\documentclass[12pt]{article} 

\usepackage{latexsym}
\usepackage{graphicx}
\usepackage{setspace}
\usepackage{fancyhdr}
\usepackage{lscape}  
\usepackage{amsmath} 
\usepackage{mathrsfs}      
\usepackage{longtable,epsfig}   
\usepackage[font=small,skip=2pt,labelfont=bf,singlelinecheck=false,justification=centering]{caption}
\usepackage{natbib}      
\usepackage{epigraph}      
\usepackage{threeparttable}        
\usepackage[colorlinks=true,pdfstartview=FitH,citecolor=blue,urlcolor=blue,linkcolor=blue]{hyperref} 
\usepackage{booktabs} 
\usepackage[font=small,labelfont=bf,singlelinecheck=false]{caption}
\usepackage{epstopdf}  
\usepackage{multirow}            
\graphicspath{{"C:/Users/mbruns/Dropbox/Non linear VARs/Drafts of paper/FiguresAndTables/"}}  
\usepackage{ragged2e}

\usepackage{amsfonts}%
\usepackage{amssymb}%

\usepackage{subcaption}    
\usepackage{lscape}

\usepackage[usenames,dvipsnames]{xcolor} 
\bibpunct{(}{)}{;}{a}{,}{,}
\def\BibTeX{{\rm B\kern-.05em{\sc i\kern-.025em b}\kern-.08em   
		T\kern-.1667em\lower.7ex\hbox{E}\kern-.125emX}}

\renewcommand\appendix{\par
	\setcounter{section}{0}%
	\setcounter{subsection}{0}%
	\setcounter{equation}{0}
	\gdef\thefigure{\Alph{section}.\arabic{figure}}%
	\gdef\thetable{\Alph{section}.\arabic{table}}%
	\gdef\thesection{\Alph{section}}%
	\gdef\thesubsection{\Alph{section}.\arabic{subsection}}%
	\gdef\theequation{\Alph{section}.\arabic{equation}}%
}
 
\usepackage{color}
\definecolor{CommentColor}{rgb}{0.953,0.371,0.258}
\definecolor{LinkColor}{rgb}{0,0.263,0.525}
\definecolor{RefColor}{rgb}{0,0.263,0.525} 
\hypersetup{colorlinks=true,%
	linkcolor=RefColor,%
	citecolor=LinkColor,%
	filecolor=LinkColor,%
	menucolor=LinkColor,%
	urlcolor=LinkColor
}    

\let\orgautoref\autoref   

\renewcommand{\autoref}{%
	\def\figureautorefname{Figure}%
	\def\tableautorefname{Table}%
	\def\sectionautorefname{Section}%
	\def\subsectionautorefname{Section}%
	\def\appendixrefname{Appendix }%
	\orgautoref}

\usepackage{xr-hyper}         
\usepackage{hyperref} 

\begin{document}
	\title{ \large{ \bf Impulse response estimation via flexible local projections} \\ - \\ Latest draft available at this \href{https://drive.google.com/file/d/1gswIrCpUcl1alCoFHFIxqG0mploEFXxf/view?usp=sharing}{link}\thanks{We are thankful to Vito Polito for helpful comments and suggestions.}} 
	
	\date{\today} 
	
	\author{Haroon Mumtaz\thanks{Queen Mary, University of London, School of Economics and Finance, Mile End road, London E1 4LJ, UK. \texttt{e-mail: \href{mailto:h.mumtaz@qmul.ac.uk}{h.mumtaz@qmul.ac.uk}}}   \and Michele Piffer\thanks{King's Business School, King's College London, Bush House, 30 Aldwych, London WC2B 4BG, UK. \texttt{e-mail: \href{mailto:m.b.piffer@gmail.com}{m.b.piffer@gmail.com}}} }
	
	
	\maketitle   
	\thispagestyle{empty}
	
	\renewcommand{\thefootnote}{\fnsymbol{footnote}} 
	\renewcommand{\thefootnote}{\arabic{footnote}} \setcounter{footnote}{0}
	\setcounter{page}{1}\pagenumbering{Roman}

	\vspace{-8mm} 
	
	\bigskip
	
	\bigskip
	
	\begin{abstract}
		This paper introduces a flexible local projection that generalises the model by \cite{Jorda2005} to a non-parametric setting using Bayesian Additive Regression Trees. Monte Carlo experiments show that our BART-LP model is able to capture non-linearities in the impulse responses. Our first application shows that the fiscal multiplier is stronger in recession than expansion only in response to contractionary fiscal shocks, but not in response to expansionary fiscal shocks. We then show that financial shocks generate effects on the economy that increase more than proportionately in the size of the shock when the shock is negative, but not when the shock is positive. 		
		
		\vspace{3mm} 
		
		
	\end{abstract}  
	
	\newpage
	\setcounter{page}{1}\pagenumbering{arabic}
	
	
	\section{Introduction}

	Estimation of impulse responses (IRFs) via local projections (LP) by \cite{Jorda2005} has become increasingly common in applied Macroeconometric analysis. A key feature of the local projection estimator is that it estimates IRFs of variable $y_{t}$ to an innovation to variable $x_{t}$ directly via linear regressions of the form $y_{t+h}=\beta_{h}x_{t}+d_{h}w_{t}+u_{t+h}$, where $w_{t}$ denotes control variables. Given their flexibility, considerable attention has been given to investigate the properties of the LP estimator, see, for instance, \cite{Stock2018} and \cite{Plagborg2021}.
	
	In their most popular specification, LP estimators impose a linearity between $y_{t+h}$ and $(x_t, w_t)$. This limitation implies that linear LPs cannot be used to study non-linear effects of the shocks of interest, for instance non-linearities on the sign or size of the shock, or on the economic conditions when the shock occurs. Some extensions of the linear LP estimator have been proposed, but they all rely on the functional form introduced to model the non-linearity. \cite{Jorda2005} proposes the use of quadratic and cubic terms. \cite{auerbach2013fiscal,Auerbach2013} and \cite{Ramey2018} use a smooth transition function and a threshold function, respectively. \cite{Ruisi2019} and \cite{Lusompa2021} use a time varying extension of LP based on parametric state-space models, while \cite{RossiTVP} provides a more general framework for modelling structural shifts. 
	
	In this paper we propose a flexible non-linear extension of the LP estimator that does not require assumptions on the functional form of the LP regression equation. We propose a non-parametric LP estimator that uses the Bayesian Additive Regression Trees (BART) model to approximate the unknown function $m_h(z_{t})$ in the more general equation $y_{t+h}=m_{h}(z_{t})+u_{t+h}$, with $z_{t}=(x_t,w_t)$. Introduced by \cite{chipman2010bart}, BART uses regression trees as its building block. Regression trees split the space of explanatory variables $z_{t}$ into sub-groups based on rules of the form $z_{t,j}<C$, where $j=1,2,..,K$. The function $m_{h}(z_{t})$ is approximated as a sum of a large number of small trees. \cite{chipman2010bart} show that BART is able to approximate highly non-linear functions accurately.  
	
	We first illustrate how BART techniques can be applied in a non-linear LP estimator, and we refer to this new methodology as BART-LP. We show that BART-LP can handle autocorrelation in the error terms, a problem already discussed in the literature of linear LP estimators. We then document the performance of BART-LP using Monte Carlo analysis. We build our simulations on three models. First, we use the same SVAR-GARCH model employed in the simulation by \cite{Jorda2005}, where a structural shock generates non-linear effects that affect the variance of the shock. Second, we use a recursive Threshold VAR model in which shocks generate stronger effects in one of the two regimes of the model. Third, we use a sign-dependent moving average model in which the true monetary policy shock generates different effects depending on the sign of the shock. In all cases, the BART-LP is capable of recovering the true impulse responses, while the linear LP typically estimates a weighted average of the true underlying non-linear impulse responses. We focus the discussion on the IRF estimation, which is separate from identification of the structural shocks. As also in \cite{Jorda2005}, we do not investigate the topic of identification in our framework.
	
	We then apply the new methodology to shed some light on two separate ongoing debates in the literature. We first apply our methodology to US fiscal spending shocks. While \cite{Auerbach2013} has argued that US fiscal multipliers are stronger in recession than in boom, their evidence has been called into question by \cite{Ramey2018}. We use our BART-LP procedure to show that the answer to this question depends on the sign of the shock. We show that the multiplier in response to a positive shock does not change significantly depending on whether the shock hits the economy in an expansion or in a recession. However, in response to a negative shock the multiplier is stronger in a recession. Compared to the models used by \cite{Auerbach2013}  and \cite{Ramey2018}, our methodology can detect both non-linearities over the state of the economy and non-linearities over the sign and size of the shock. Our results hence help reconcile the difference found by \cite{Auerbach2013}  and \cite{Ramey2018}, who used models that are more suitable to detect non-linearities over the state of the economy, but not over the sign of the shocks. Last, we revisit the analysis by \cite{forni2022nonlinear} on financial shocks. We confirm their result that negative financial shocks generate detrimental effects on output that increase more than proportionally in the size of the shock. These effects cannot otherwise be detected by a linear model, which is doomed to underestimate the effect of strong negative financial shocks. 
	
	This paper relates to the literature that studies how BART techniques can be used in Macroeconometrics. \cite{huber2021inference} introduce a VAR model where the dynamics of the endogenous variables are modelled using BART. The authors model the impact of uncertainty shocks using their proposed model. \cite{HUBER2020} extend the BART-VAR to a mixed frequency setting and evaluate the forecasting performance of the model. \cite{clark2021tail} show that multivariate BART regression models perform well in terms of tail forecasting. To the best of our knowledge, our paper is the first one to use BART in an LP framework. The paper is also part of a broad literature that studies the advantages of IRF estimation using LP estimators, relative to constructing IRFs on vector autoregressive models. Several contributions document the performance of LP estimators, including \cite{gonccalves2021when}, \cite{kilian2011reliable}, \cite{alloza2019dynamic}, \cite{breitung2019projection} \cite{herbst2021bias},  and \cite{bruns2022comparison}. While LP estimators are usually proposed in a frequentist setting, we follow \cite{miranda2021bayesian} and take a Bayesian approach to LP, yet in a non-linear framework. 
	
	The paper is organised as follows. \autoref{sec_model} presents the empirical model. \autoref{sec_MonteCarlo} reports the results from the simulation exercise. \autoref{empirics} shows the application to fiscal and financial shocks. \autoref{sec_conclusions} concludes.

	\section{Flexible local projections} \label{sec_model}
	
	In this section we outline the methodology, which we refer to as the BART-LP model, or flexible local projections. We discuss how BART-LP approximates the unknown conditional expectation function of local projection models, discuss the prior, and outline the posterior sampler. We then discuss how to compute generalized impulse responses to structural shocks within our framework. 	
	
	\subsection{The BART approximation}	
	
	We work with the equation 
	\begin{equation}
		y_{t+h}=m_{h}\big(x_{t},\boldsymbol{z}_{t}, \boldsymbol{w}_{t+h}^{(h)}\big)+ \epsilon_{t+h}^{(h)}, \label{FLP}
	\end{equation} where $y_{t+h}$ denotes the scalar variable of interest, $h=0,1,...,H$ is the impulse response horizon, and $\epsilon_{t+h}^{(h)}$ satisfies $E\big(\epsilon_{t+h}^{(h)} | x_{t},\boldsymbol{z}_{t}, \boldsymbol{w}_{t+h}^{(h)}\big)=0$. We aim to study how $y_{t+h}$ responds to a change in the scalar variable $x_{t} $. The vector $\boldsymbol{z}_{t}$ contains observable control variables, possibly including lagged values of $y_{t}$ and contemporaneous and/or lagged values of other variables. The vector $\boldsymbol{w}_{t+h}^{(h)}$ contains additional control variables in the form of estimated residuals, as explained in \autoref{sec_autocorrelation}. While $(x_t, \boldsymbol{z}_{t})$ are the same for every regression model $h$, $\boldsymbol{w}_{t+h}^{(h)}$ can potentially change. The function $m_h(.)$ captures the true unknown conditional expectation function. The residual $\epsilon^{(h)}_{t+h}$ is assumed to be normally distributed with variance $\sigma_{t+h}^{2 ~(h)}$. As noted in \cite{Jorda2005}, $\epsilon^{(h)}_{t+h}$ is serially correlated for $h\ge1$.
	
	It is common in the literature to assume a functional form for $m_{h}\big(x_{t},\boldsymbol{z}_{t}, \boldsymbol{w}_{t+h}^{(h)}\big)$. The most popular applications of LPs use a function of the type
	\begin{align}
		y_{t+h}& = ~~~~~~~ g(q_{t}) \big[ \alpha^{(h)}_0 x_{t} + \boldsymbol{\alpha}_1'^{(h)} \boldsymbol{z}_{t} + \boldsymbol{\alpha}_2'^{(h)} \boldsymbol{w}_{t+h}^{(h)} \big] + \\
		& + \big(1- g(q_{t}) \big) \big[ \beta^{(h)}_0 x_{t} + \boldsymbol{\beta}_1'^{(h)} \boldsymbol{z}_{t} + \boldsymbol{\beta}_2'^{(h)} \boldsymbol{w}_{t+h}^{(h)} \big] + \epsilon_{t+h}^{(h)}.
	\end{align} The special case of a linear model sets $g(q_{t})=1$, $\forall ~ t$, and estimates the impulse response to a shock to $x_t$ using the estimates for $\{\alpha^{(h)}_0\}_{h=0}^H$ (\citealp{Jorda2005}). Non-linear applications usually specify a transition variable $q_t$, assume a specific non-linear functional form for $g(.)$, and compute non-linear impulse responses as a function of the estimates for $\{\alpha^{(h)}_0, \beta^{(h)}_0 \}_{h=0}^H$. For example, the smooth transition formulation by  \cite{auerbach2013fiscal,Auerbach2013} sets $g(.)$ equal to the logistic function, while the threshold formulation by \cite{Ramey2018} and \cite{alpanda2021state} sets $g(.)$ equal to the indicator function. 
	
	Our paper differs from the existing literature by approximating the unknown conditional expectation function $m_{h}\big(x_{t},\boldsymbol{z}_{t}, \boldsymbol{w}_{t+h}^{(h)}\big)$ using Bayesian Additive Regression Trees (BART). It is assumed that
	\begin{equation}
		m_{h}( x_{t},\boldsymbol{z}_{t}, \boldsymbol{w}_{t+h}^{(h)})  \approx f_{h}(x_{t},\boldsymbol{z}_{t}, \boldsymbol{w}_{t+h}^{(h)}|\Gamma^{(h)},\boldsymbol{\mu}^{(h)}) = \sum\limits_{j=1}^{J} f_{h,j}(x_{t},\boldsymbol{z}_{t}, \boldsymbol{w}_{t+h}^{(h)}|\Gamma_{j}^{(h)},\boldsymbol{\mu}_{j}^{(h)}), \label{bart1}
	\end{equation} where $f_{h,j}( x_{t},\boldsymbol{z}_{t}, \boldsymbol{w}_{t+h}^{(h)}|\Gamma^{(h)}_j,\boldsymbol{\mu}_j^{(h)})$ denotes a single regression tree $j$ at horizon $h$ and $f_h(x_{t},\boldsymbol{z}_{t}, \boldsymbol{w}_{t+h}^{(h)}|\Gamma^{(h)},\boldsymbol{\mu}^{(h)})$ denotes the sum of $J$ regression trees. The parameters of the regression trees are the tree structures $\Gamma^{(h)} = [\Gamma^{(h)}_1, .., \Gamma^{(h)}_J]$ and the terminal nodes (or leaves) $\boldsymbol{\mu}^{(h)} = (\boldsymbol{\mu}_{1}^{(h)'}, .., \boldsymbol{\mu}_{J}^{(h)'})'$, with $\boldsymbol{\mu}_{j}^{(h)}$ of dimension $B_j \times 1$ and $B_j$ the number of terminal nodes of tree $j$. As an illustration, each regression tree divides the space of each explanatory variable by using binary splitting rules. Denoting $X_{i}$ as the \textit{i-th} entry of the vector $\big(x_{t},\boldsymbol{z}_{t}, \boldsymbol{w}_{t+h}^{(h)}\big)$, these rules are defined as:
	\begin{align}
		X_{i}  & \leq c, \label{rules}\\
		X_{i}  & > c, \nonumber
	\end{align} with $c$ the threshold value. Observations are assigned according to these splitting rules, and the terminal nodes return the fitted value conditional on the split. The fitted value of the dependent variable, based on a single regression tree, is then given by%
	\begin{equation} 
		f_{h,j}(x_{t},\boldsymbol{z}_{t}, \boldsymbol{w}_{t+h}^{(h)}|\Gamma_j^{(h)},\boldsymbol{\mu}_j^{(h)})  =\sum_{b=1}^{B_j} I(X_{i})  \mu_{j,b}^{(h)},
	\end{equation} where $I(.)$ denotes an indicator function that equals 1 if $X_{i}$ belongs to the set defined by the splitting rule implicit in $\Gamma_j^{(h)}$. Note that the complexity of each tree is determined by $B_j$, the number of terminal nodes. We refer the reader to the Online Appendix for an illustrative example of a regression tree, and to \cite{Hill2020} for a comprehensive review.
	
	The model in equation \eqref{bart1} approximates $m_{h}\big(x_{t},\boldsymbol{z}_{t}, \boldsymbol{w}_{t+h}^{(h)}\big)$ using a sum $J$ trees. Each tree in the sum is restricted to be small \textit{a priori} to avoid overfitting, and thus explains a small proportion of $y_{t+h}$ and is a `weak learner'. \cite{chipman2010bart} show that a low value of $J$ reduces predictive accuracy. As $J$ increases, predictive performance initially improves, but this improvement tapers off, eventually. In practice, studies such as \cite{HUBER2020} note that the difference in predictive accuracy is negligible for $J>150$ and complex functions can be easily approximated using $J=200$ or $250$.
	
	The BART approximation of the relationship between $y_{t+h}$ and $x_{t}$ has implications for the properties of the impulse responses. As the regression trees split the space of the covariates via rules of the type shown in equation \eqref{rules}, the estimated predictions on the right-hand side of equation \eqref{IRF} are dependent on their history. Similarly, the shock $d$ to variable $x_t$ can lead to predictions that proportionally differ if the size and sign of the shock leads to the covariate space where the relationship between $y_{t+h}$ and $x_{t}$ is substantially different from the `average' impact. We now discuss estimation, and then return to a detailed discussion of non-linear impulse responses in the BART-LP model in \autoref{sec_irfs}. 
	
	\subsection{Estimation} \label{sec_estimation}
	
	The model in equation \eqref{FLP} can be estimated using the MCMC algorithm described in \cite{chipman2010bart}, which we summarize here for completeness.
	
	\subsubsection{Priors}
	
	The prior distributions proposed by \cite{chipman2010bart} play a crucial role, as they are devised to reduce the possibility of overfitting. The joint prior for the parameters of the $J$ trees of the model at each horizon $h$ is factored as follows:
	\begin{equation}
		p\big((\Gamma_{1}^{(h)},\boldsymbol{\mu}_{1}^{(h)}),(\Gamma_{2}^{(h)},\boldsymbol{\mu}_{2}^{(h)}),...,(\Gamma_{J}^{(h)},\boldsymbol{\mu}_{J}^{(h)})\big)  = \prod_{j=1}^{J} p(\boldsymbol{\mu}_{j}^{(h)}|\Gamma_{j}^{(h)})p(\Gamma^{(h)}_{j}),
	\end{equation} where $p(\boldsymbol{\mu}^{(h)}_{j}|\Gamma^{(h)}_{j})=\prod_{b=1}^{B_j} p(\mu_{b,j}^{(h)}|\Gamma_{j}^{(h)})$.
	
	The prior for the tree structure $\Gamma_{j}^{(h)}$ depends on the probability that the node at depth $d=0,1,2,..$ is not a terminal node. This prior probability is given by $\alpha(1+d)^{-\beta}$ where $\alpha\in(0,1)$ and $\beta>0$. Higher values of $\beta$ and smaller values of $\alpha$ reduce this probability and impose a stronger belief that the tree has a simple (i.e. shorter) structure. We follow the recommendation by \cite{chipman2010bart} and set $\alpha=0.95$ and $\beta=2$. The prior for the threshold value $c$ implies that this parameter is assumed to be uniform over the range of the values taken by the variables. In the default setting, the choice of splitting variable is also assumed to be uniform across the regressors.
	
	To define $p(\boldsymbol{\mu}^{(h)}_{j}|\Gamma^{(h)}_{j})$ \cite{chipman2010bart} first transform the dependent variable so that it lies between $-0.5$ and $0.5$. As a consequence, $m_{h}\big(  x_{t},\boldsymbol{z}_{t}, \boldsymbol{w}_{t+h}^{(h)} \big)  $ is also expected to lie between these values. The prior $p(\boldsymbol{\mu}^{(h)}_{j}|\Gamma^{(h)}_{j})$ is assumed to be normal $N(0,S)  $. The variance $S$ is set as $\frac{1}{2\kappa(J^{0.5})}$, with $\kappa$ set to 2, the value recommended by \cite{chipman2010bart}. Under this default prior, there is a 95\% probability that the conditional mean of the dependent variable lies between $-0.5$ and $0.5$.
	
	A conjugate inverse $\chi^{2}$ prior is used for the variance $\sigma_{t+h}^{2 ~(h)}$. The hyperparameters of the prior distribution are set by using an estimate $\hat{\sigma}_{t+s}^{2~(h)}$ of the variance obtained from a linear regression. If the true model is non-linear $\hat{\sigma}_{t+s}^{2 ~(h)}$ will be biased upwards. Under the default prior, the hyperparameters are chosen so that $\Pr(  \sigma_{t+s}^{(h)}<\hat{\sigma}_{t+s}^{(h)})  =0.9$.
	
	The total number of trees $J$ is fixed.
	
	\subsubsection{MCMC algorithm}

	The MCMC algorithm devised by \cite{chipman2010bart} samples from the conditional posterior distributions of $\sigma_{t+h}^{2~(h)}$ and the parameters of the regression trees in each iteration.\footnote{Intuitive descriptions of this MCMC algorithm can be found in \cite{clark2021tail} and \cite{Hill2020}.} Each iteration of the algorithm samples from the following conditional posteriors: 
	
	\begin{enumerate}
		\item conditional on the trees, the error variance can be easily drawn from the inverse Gamma distribution;
		
		\item the conditional posterior distribution of the tree structure is not known in closed form and a Metropolis-Hastings algorithm is used. Define $R^{(h)}_{j}$ as the residual:
		\begin{equation}	
			R^{(h)}_{j}=y_{t+h}-\sum_{i\neq j} f(x_{t},\boldsymbol{z}_{t}, \boldsymbol{w}_{t+h}^{(h)}|\Gamma^{(h)}_{j},\boldsymbol{\mu}^{(h)}_{j}).
		\end{equation} The $j-th$ tree is proposed using the density $q(  \Gamma_{j}^{new}, \Gamma_{j}^{old})  $. \cite{chipman2010bart} use a proposal density that incorporates 4 moves: (i) splitting the node into two new nodes (grow), (ii) transforming adjacent nodes to terminal node (prune), (iii) changing the decision rule of an interior node (change), (iv) swapping a decision rule between a node that is above and the node before it (swap). The probabilities associated with these moves are fixed at $0.25,0.25,0.4$ and $0.1$ respectively. The proposed tree structure $\Gamma_{j}^{new}$ is accepted with probability
		\begin{equation}
			\alpha=\frac{q(\Gamma_{j}^{new},\Gamma_{j}^{old})  p(R_{j}^{(h)}|\Gamma_{j}^{new},\sigma_{t+h}^{2~h}) p(\Gamma_{j}^{new})}{q(\Gamma_{j}^{old},\Gamma_{j}^{new})  p(R_{j}^{(h)}|\Gamma_{j}^{old},\sigma_{t+h}^{2})  p(\Gamma_{j}^{old})},
		\end{equation} where $p(R_{j}|\Gamma_{j},\sigma_{t+h}^{2~h})$ is the conditional likelihood and $p(  \Gamma_{j})  $ denotes the prior. This step is repeated for $j=1,2,..,J$ trees;
		
		\item the conditional posterior distribution of the terminal node parameters is Gaussian with the parameters known in closed form. Therefore, the draw of $\boldsymbol{\mu}^{(h)}_{j}$ for $j=1,2,..,J$ can be carried out in a straightforward manner;
		 
		\item given a draw of the model parameters conditioning on $(x_{t},\boldsymbol{z}_{t}, \boldsymbol{w}_{t+h}^{(h)})$, the predicted value can be computed as  
		\begin{equation}
		E\left(  y_{t+h}|x_{t},\boldsymbol{z}_{t}, \boldsymbol{w}_{t+h}^{(h)}\right)=\sum_{j=1}^{J}f_h(x_{t},\boldsymbol{z}_{t}, \boldsymbol{w}_{t+h}^{(h)}|\Gamma^{(h)}_{j},\boldsymbol{\mu}^{(h)}_{j}) = \sum_{j=1}^{J}\sum_{b=1}^{B_j} I(x_{t},\boldsymbol{z}_{t}, \boldsymbol{w}_{t+h}^{(h)}) \mu_{j,b}^{(h)},
		\end{equation} with $I(x_{t},\boldsymbol{z}_{t}, \boldsymbol{w}_{t+h}^{(h)})$ an indicator function equal to 1 if $(x_{t},\boldsymbol{z}_{t}, \boldsymbol{w}_{t+h}^{(h)})$ belongs to the splitting rule implied by $[\Gamma_1^{(h)}, ..., \Gamma_J^{(h)}]$.
	\end{enumerate}

	\subsection{Autocorrelation} \label{sec_autocorrelation}
	
	The residual term in LP models is known to be autocorrelated, a feature that must be taken into account in the estimation. In the case of linear local projections, it has been shown that the residual at horizon $h$ follows a $MA(h-1)$ process, see for instance \cite{Lusompa2021}. \cite{Lusompa2021} suggests a GLS procedure whereby the autocorrelation is eliminated by including leads of the LP residuals from horizon $h=0$ in the conditioning set.\footnote{\cite{Lusompa2021} 	suggests an efficient strategy that transforms the dependent variable of the LP regressions and does not require  one to explicitly include the horizon $0$ residuals as regressors.} 
	
	The non-parametric setting considered in this paper encompasses non-linear models. For the purpose of illustration, consider a simple non-parametric AR(1) model
	\begin{equation}
		y_{t+1}=v(y_{t};A_{1})+e_{t+1}.
	\end{equation} Iterating the process forward 3 periods as an example gives
	\begin{align*}
		y_{t+2}  & =v\big(v(y_{t};A_{1})+e_{t+1};A_{1}\big)  +e_{t+2},\\
		y_{t+3}  & =v\big(v(  v(y_{t};A_{1})+e_{t+1};A_{1}\big)	+ e_{t+2};A_{1})  +e_{t+3}.
	\end{align*} It is useful to compare this with a BART-LP for this horizon:
	\begin{equation}
		y_{t+3}=f_{3}(y_{t}| \Gamma_3, \boldsymbol{\mu}_3)+\epsilon_{t+3}^{(3)}.
	\end{equation} The function $f_3(y_{t}| \Gamma_3, \boldsymbol{\mu}_3)$ approximates the non-linear relationship between $y_{t}$ and its lead, but does not account for the dependence between the dependent variable and lagged shocks. Thus in this setting, the residual $\epsilon_{t+3}^{(3)}$ is a non-linear function of $e_{t+1}$ and $e_{t+2}$, and has a non-linear autocorrelation structure.
	
	In general, the Voltera expansion of any non-linear time-series shows its complex dependence on past shocks:%
	\begin{equation}
		y_{t}= 
		\sum\limits_{i=0}^{\infty}
		\phi^{i}e_{t-i}+ \sum\limits_{i=0}^{\infty}
		\sum\limits_{i=j}^{\infty}
		\zeta_{ij}e_{t-i}e_{t-j}+\sum\limits_{i=0}^{\infty}
		\sum\limits_{i=0}^{\infty}
		\sum\limits_{k=0}^{\infty}
		\phi_{ij}e_{t-i}e_{t-j}e_{t-k}+...
	\end{equation} To account for this autocorrelation, we propose to include an estimate of the shocks, $\boldsymbol{w}_{t+h}^{(h)}=(\hat{e}_{t+1},\hat{e}_{t+2,}...,\hat{e}_{t+h-1})$, as additional covariates in the $h-$period BART-LP. Following \cite{Lusompa2021} we construct $\boldsymbol{w}_{t+h}^{(h)}$ at every horizon $h \ge 1$ using the residuals of the period $h=0$ flexible local projection $y_{t}=f_{0}(x_t, \boldsymbol{z}_{t}, \boldsymbol{w}_{t}^{(0)}|\Gamma^{(0)}, \boldsymbol{\mu}^{(0)}) + \epsilon_{t}^{(0)}$, with $\boldsymbol{w}_{t}^{(0)} = \boldsymbol{0}$. Then, the flexible local projection for period $h$ is specified as
	\begin{equation}
		y_{t+h}=f_{h}(x_{t},\boldsymbol{z}_{t},\boldsymbol{w}_{t+h}^{(h)}|\Gamma^{(h)},\boldsymbol{\mu}^{(h)})  +\epsilon^{(h)}_{t+h}.	
	\end{equation} The BART approximation of the true non-linear function $m_{h}(x_{t},\boldsymbol{z}_{t},\hat{w}_{t+h}^{(h)})$ proxies the non-linear dependence of $y_{t+h}$ on $e_{t+1},e_{t+2,}...,e_{t+h-1}$ and ameliorates the autocorrelation in $\epsilon_{t+h}^{(h)}$.

	\subsection{Generalized structural impulse responses} \label{sec_irfs}
	
	The computation of impulse responses to structural shocks typically relies on two conceptually different pillars: an estimation procedure for the impulse responses, and an identification scheme for the structural shock of interest. Our paper only aims to advance the literature on the former, and does not investigate the topic of identification in a non-parametric framework. 
	
	Define $\boldsymbol{y}_t$ a $k \times 1$ vector of variables of interest, with $x_t$ one of the entries of $\boldsymbol{y}_t$. Define $\phi_{i,h}$ the impulse response of variable $y_{i,t}$ to a structural shock to variable $x_t$ of size $d$, with $i=1,..,k$. Following \cite{koop1996impulse}, $\phi_{i,h}$  is given by 
	\begin{equation} 
		\phi_{i,h}=E ( y_{i,t+h}|x_{t}=d;\boldsymbol{n}_{t} )  -E(  y_{i,t+h}|x_{t},\boldsymbol{n}_{t}). \label{IRF}
	\end{equation} We compute generalized impulse responses numerically using the following algorithm:
\begin{enumerate} 
	\item estimate the model for $h=0$,
	\begin{equation}
		y_{i,t} = f_0\big(x_t, \boldsymbol{z}_{t}, \boldsymbol{w}_{t}^{(0)}|\Gamma^{(0)},\boldsymbol{\mu}^{(0)} \big) + \epsilon_{i,t}^{(0)} , 
	\end{equation} with $\boldsymbol{w}_{t}^{(0)} = \boldsymbol{0}$, and store $D$ vectors of dimension $T \times 1$ containing the estimated residuals associated with $D$ posterior draws, $\{ \hat{\epsilon}_{i,t}^{(0),d}\}_{t=1}^T$, $d = 1, .., D$;
	\item for the generic draw $d$ of the residuals $\{ \hat{\epsilon}_{i,t}^{(0),d}\}_{t=1}^T$ and for the generic horizon $h$, estimate the model
		\begin{equation}
		y_{i,t+h} = f_h \big(x_t, \boldsymbol{z}_{t}, \boldsymbol{w}_{t+h}^{(h)} | \Gamma^{(h)},\boldsymbol{\mu}^{(h)} \big) + \epsilon_{i,t+h}^{(h)}, \label{eq_temp1}
	\end{equation} with
\begin{equation} 
\boldsymbol{w}_{t+h}^{(h)} = (\hat{\epsilon}_{i,t}^{(0),d}, \hat{\epsilon}_{i,t-1}^{(0),d}, ..., \hat{\epsilon}_{i,t-h+1}^{(0),d});
\end{equation} 
	\item compute the predicted values $(\hat{y}_{i,h}^{0,d},~ \hat{y}_{i,h}^{1,d})$ associated with one posterior draw from model \eqref{eq_temp1} conditioning on 
	\begin{align}
		\big(x_t = \bar{x},~~~~~~ &\boldsymbol{z}_t = \bar{\boldsymbol{z}}, ~\boldsymbol{w}_{t+h}^{(h)} = \bar{\boldsymbol{w}}^{(h)} \big), \label{eq_temp3} \\
	\big(x_t = \bar{x}+d, ~&\boldsymbol{z}_t = \bar{\boldsymbol{z}}, ~ \boldsymbol{w}_{t+h}^{(h)} = \bar{\boldsymbol{w}}^{(h)} \big),  \label{eq_temp4}
	\end{align} respectively, with $\big(\bar{x}, \bar{\boldsymbol{z}}, \bar{\boldsymbol{w}}^{(h)} \big)$ defined below. Compute $\psi_{i,h}^{d} = \hat{y}_{i,h}^{1,d} - \hat{y}_{i,h}^{0,d}$ and store
		\begin{equation}
		\big( \hat{y}_{i,h}^{0,d}, ~~\hat{y}_{i,h}^{1,d}, ~~ \psi_{i,h}^d \big) ;
	\end{equation}
	\item repeat steps 2-3 for $h=0, .., H$, store $\{\big( \hat{y}_{i,h}^{0,d}, ~~\hat{y}_{i,h}^{1,d}, ~~ \psi_{i,h}^d \big) \}_{h=0}^H$ given the same posterior draw $d$ for $\{\hat{\epsilon}_{i,t}^{(0),d}\}_{t=1}^T$;
	\item repeat steps 2-4 for $d=1, ..,  D$;
	\item compute the average across posterior draws
		\begin{equation}
		\psi_{i,h} = \frac{1}{D}\sum_{d=1}^{D} \hat{y}_{i,h}^{1,d} - \frac{1}{D}\sum_{d=1}^{D} \hat{y}_{i,h}^{0,d} \label{eq_temp2}.
	\end{equation}
\end{enumerate} 	
	
The exact implementation of the above procedure depends on how the structural shocks are identified. Different options are in principle available. One option is to follow \cite{Plagborg2021} and \cite{Barnichon2019} and replicate a recursive identification scheme by using appropriate control variables. For example, if the aim is to estimate the response of GDP to an interest rate shock that is restricted to have a zero contemporaneous impact on GDP and CPI in a trivariate model, one can set $x_t$ equal to the policy interest rate and add contemporaneous GDP and CPI into $\boldsymbol{z}_{t}$, which will also include $L$ lags of the variables. Alternatively, one can set $x_t$ equal to either a proxy for the structural shock of interest, or equal to the true realizations of the shocks, if available, and set $\boldsymbol{z}_t = (\boldsymbol{y}_{t-1}', .., \boldsymbol{y}_{t-L}')'$. Either way, the generalized impulse responses are computed as $\phi_{i,h} = \psi_{i,h}$, with $\psi_{i,h}$ from equation \eqref{eq_temp2}. 

An alternative approach is to follow \cite{Jorda2005} more closely and use a separate impulse vector. One can estimate a preliminary SVAR on $\boldsymbol{y}_t$ and use a preferred identification approach to estimate the impulse vector $\boldsymbol{d}$ of variables $\boldsymbol{y}_t$ to a shock to variable $x_t$ of size $d$. The above algorithm can then be run by setting $x_t=0$ and $\boldsymbol{z}_t = (\boldsymbol{y}_{t-1}', .., \boldsymbol{y}_{t-L}')'$. Step 1 of the algorithm is still required to generate $\{ \hat{\epsilon}_{i,t}^{(0),d}\}_{t=1}^T$, $d = 1, .., D$. Steps 2 to 4 are then run for $h=1,2,..,H$, replacing equations \eqref{eq_temp3}-\eqref{eq_temp4} with
	\begin{align}
	\big(x_t = 0,~ &\boldsymbol{z}_t = \bar{\boldsymbol{z}}, ~~~~~~\boldsymbol{w}_{t+h}^{(h)} = \bar{\boldsymbol{w}}^{(h)} \big), \\
	\big(x_t = 0, ~&\boldsymbol{z}_t = \bar{\boldsymbol{z}} + \boldsymbol{d}, ~ \boldsymbol{w}_{t+h}^{(h)} = \bar{\boldsymbol{w}}^{(h)} \big).
\end{align} The generalized impulse responses are then computed as $\phi_{i,h} = d_i$ for $h=0$ with $d_i$ the $i-th$ entry of $\boldsymbol{d}$, and as $\phi_{i,h} = \psi_{i,h}$ from \eqref{eq_temp2} for $h=1,..,H$.

Last, to be operational, the above procedure requires specifying the conditioning values $\big( \bar{x}, \bar{\boldsymbol{z}}, \bar{\boldsymbol{w}}^{(h)} \big)$ as well as the shock $d$. The former can be set differently depending on whether the shock is simulated to hit the economy at any point in time or on a subset of periods. Our baseline specification sets $\big( \bar{x}, \bar{\boldsymbol{z}}, \bar{\boldsymbol{w}}^{(h)} \big)$ equal to the sample average, which is computed within each regression model $h$. $d$, instead, can be set in accordance to the sign and size of the intended shock.  

\section{Monte Carlo simulation} \label{sec_MonteCarlo}

We use Monte Carlo simulations to assess if BART-LP can recover the true non-linear patters of a data generating process. We use three different data generating processes: a SVAR-GARCH model, a Threshold VAR model, and a sign-dependent Moving Average model.

\subsection{SVAR-GARCH model}

The first model we use for simulations is the model from Section III.B in \cite{Jorda2005},
\begin{align}
	\begin{pmatrix}
		y_{1t} \\ y_{2t} \\ y_{3t}  
	\end{pmatrix} &= A \begin{pmatrix}
		y_{1t-1} \\ y_{2t-1} \\ y_{3t-1} 
	\end{pmatrix} + B h_{t} + \begin{pmatrix}
		\sqrt{h_t} \epsilon_{1t} \\ \epsilon_{2t} \\ \epsilon_{3t}
	\end{pmatrix}, \\
	(\epsilon_{1t}, \epsilon_{2t}, \epsilon_{3t})' &= \boldsymbol{\epsilon}_t \sim N(\boldsymbol{0},I_3), \\
	h_t &= 0.5 + 0.5 h_{t-1} + 0.3 \sqrt{h_t}\epsilon_{1t}, \\
	B = \begin{pmatrix}
		-1.75 \\ -1.5 \\ 1.75
	\end{pmatrix}~~~&~~~~~  A  =\begin{pmatrix}
		0.5 & -0.25 & 0.25 \\ 0.75 & 0.25 & 0.25 \\ -0.25 & -0.25 & 0.75
	\end{pmatrix}.
\end{align} In this model, each shock $i$ affects only variable $i$ contemporaneously, while all shocks affect all variables after one period. Contrary to the second and the third shock, the first shock features time-varying variance.\footnote{We code the simulation exercise following the exact code available in the replication files from \cite{Jorda2005}.} 

\begin{figure}[h!!]    
	\centering 
	\caption{Monte Carlo simulation - SVAR-GARCH} 
	\includegraphics[viewport = 110 320 490 525, scale = 0.93]{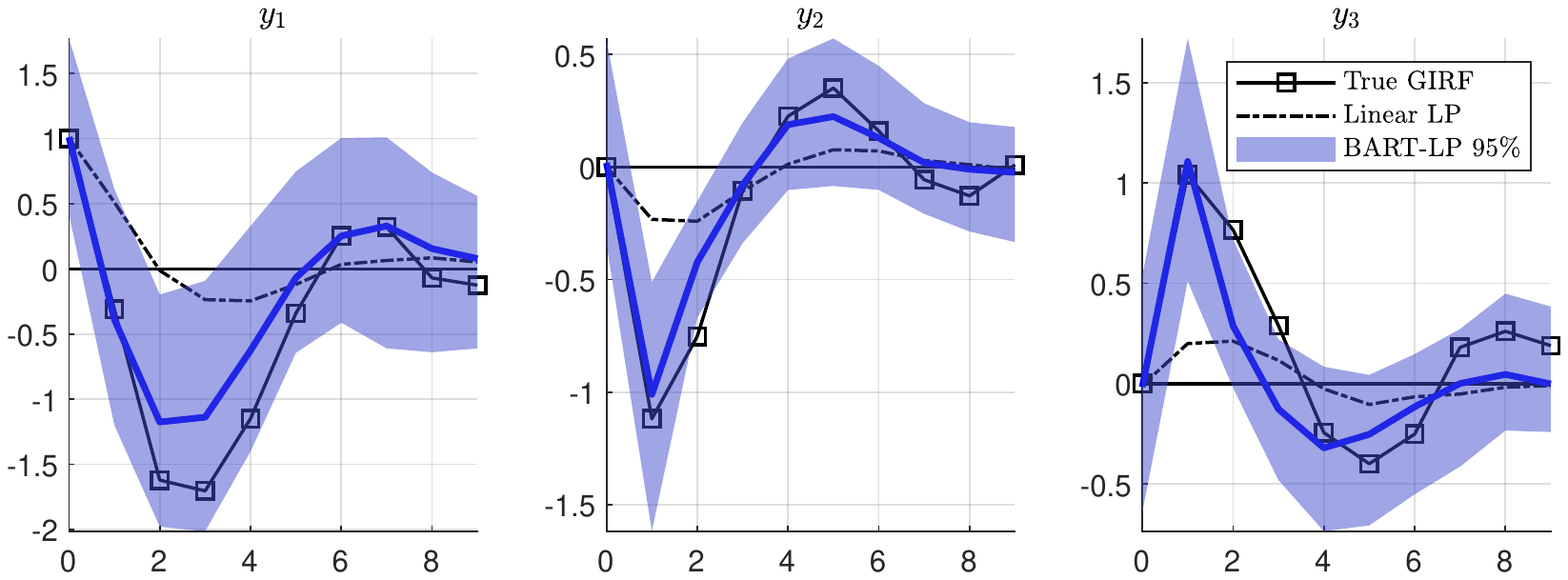} 
	\label{fig:MC_SVAR_GARCH_7_T200}   
\end{figure}  

We use a simulation exercise to study how well the BART-LP methodology recovers the true impulse responses associated with a shock to $y_{1t}$ of size $1$. We first use generalized impulse responses to simulate the true impulse responses associated with $\epsilon_{1t}=1$. We then generate simulated data for 300 periods and discard the first 100, as in \cite{Jorda2005}. The remaining 200 observations are used to estimate the pointwise linear LP impulse response, as well as the impulse response computed from the BART-LP method, setting $d=1$. We replicate the exercise 100 times and store the generated estimates. The estimated model uses 2 lags, and sets $x_t$ equal to the true realizations of $\epsilon_{1t}$. We set $J=250$.

\autoref{fig:MC_SVAR_GARCH_7_T200} reports the results of the analysis. The squared black line shows the true generalized impulse responses. The shock increases $y_{1t}$ on impact by 1, and generates no contemporaneous response in $(y_{2t},y_{3t})$. It subsequently generates an oscillating pattern in $(y_{1t}, y_{2t})$ and a hump-shaped response of $y_{3t}$. The dashed line shows the pointwise mean impulse response from the linear LP model, with the mean computed on the pointwise estimates along the 100 iterations. The continuous line and the shaded area report the pointwise median and 95\% band computed over the 100 median responses from the BART-LP method, where within each iteration, the median response from BART-LP is computed over 2,000 posterior draws. We see that BART-LP does remarkably well in replicating the response of all variables, capturing both the timing and the magnitude of the response correctly. By contrast, the response estimated with the linear model estimated an attenuated effect, underestimating the effect of the shock in the short horizon.

\subsection{Threshold VAR model} 
The second model we use for simulations is 
\begin{subequations}
	\begin{align}
		\boldsymbol{y}_t &= [\Pi_1 \boldsymbol{y}_{t-1} + B_1 \boldsymbol{\epsilon}_t] \cdot \text{I}(y_{3,t-1} \le 0) ~+ \\
		&~ +[\Pi_2 \boldsymbol{y}_{t-1} + B_2 \boldsymbol{\epsilon}_t] \cdot \text{I}(y_{3,t-1} > 0), \\
		\boldsymbol{\epsilon}_t& \sim N(\boldsymbol{0}, I), 
	\end{align} \label{eq_TVAR}
\end{subequations} with $\boldsymbol{y}_t = (y_{1t}, y_{2t}, y_{3t})'$, $\boldsymbol{\epsilon}_t = (\epsilon_{1t}, \epsilon_{2t}, \epsilon_{3t})'$ and
\begin{align}
	\Pi_1 & =\begin{pmatrix}
		0.25 & 0.25 & -0.25 \\ 
		-0.25 & 0.25 & -0.25 \\ 
		0.25 & 0.25 & 0.15
	\end{pmatrix}, ~~~~~~ B_1 = \begin{pmatrix}
		0.10 &  0  & 0 \\
		-0.20 &   0.15 & 0 \\
		0.10 &  -0.10 &  1  
	\end{pmatrix}, \\
	\Pi_2 & =\begin{pmatrix}
		0.50 & 1.25 & -1.75 \\  
		-0.25 & 0.50 & -1.25 \\ 
		0.25 & 0.25 & 0.15 
	\end{pmatrix}, ~~~~~~ B_2 = \begin{pmatrix}
		0.10 &  0  & 0 \\
		-0.20 &   0.15 & 0 \\
		0.10 &  -0.10 &  0.40
	\end{pmatrix}.
\end{align} The model is a recursive Threshold Vector Autoregressive model that jumps across two regimes depending on the endogenous evolution of the third variable. See, for instance, \cite{castelnuovo2018uncertainty}.

We first compute the true generalized impulse responses to a positive one-standard-deviation shock to a $y_{3t}$ in regime 1 by setting the initial condition of the impulse response to $\boldsymbol{y}_0 = \boldsymbol{0}$. We then study how well the linear-LP and the BART-LP estimators recover the true impulse responses. We generate a dataset of 300 observations, discard the first 100 and use the remaining 200 to estimate the impulse responses. We compute the point estimates of the linear LP, and compute the pointwise median over 2,000 posterior draws from the BART-LP estimates. We then repeat the exercise over 100 iterations. The estimated models use 4 lags and set $(x_t, \boldsymbol{z}_t)$ to replicate the recursive ordering within LP models, as discussed in \autoref{sec_irfs}.  We set $J=250$.

\begin{figure}[hh!!!]     
	\centering 
	\caption{Monte Carlo simulation - TVAR}  
	\includegraphics[viewport = 110 320 490 525, scale = 0.93]{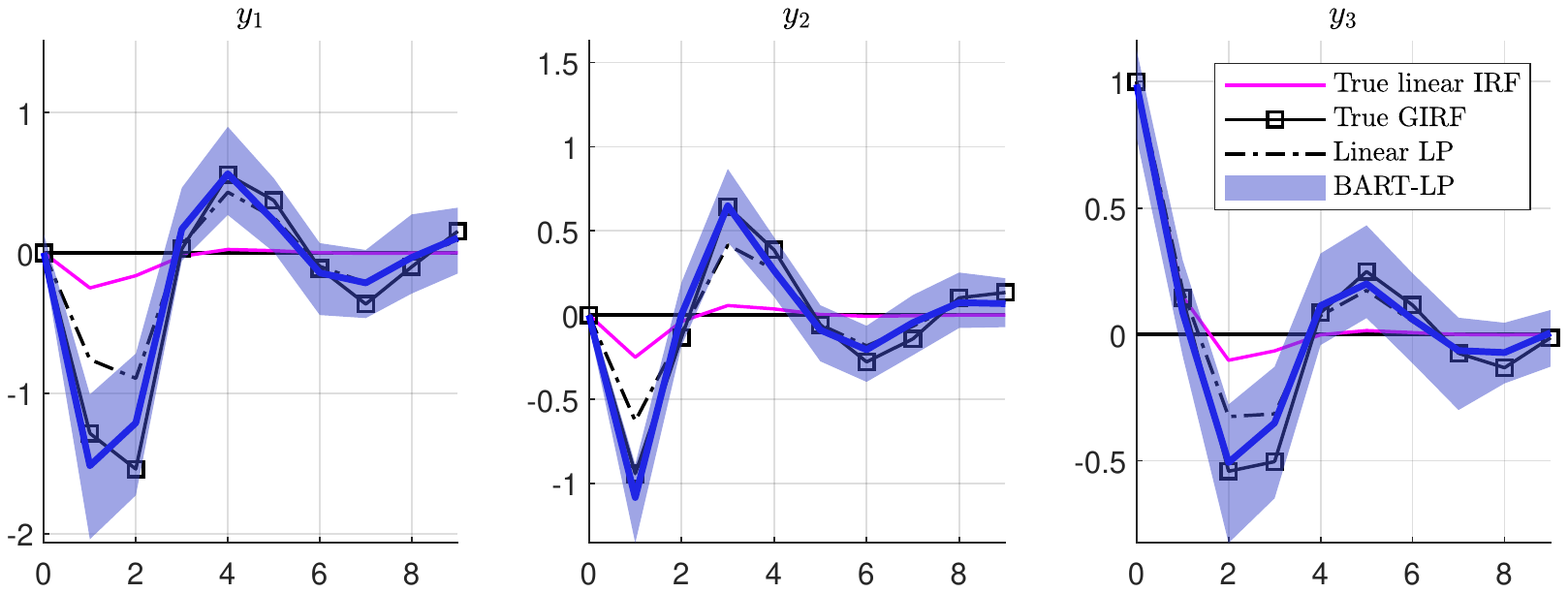} 
	\label{fig:MC_TVAR_133_cutoff3_shockto3_T200}   
\end{figure} 

The shock hits the system when the system is in regime 1. Under linearity, the model would stay in regime 1, with an impulse response uniquely pinned down by $(\Pi_1,B_1)$. Instead, the non-linearity of the model implies an endogenous evolution across regimes, in accordance with the endogenous response of $y_{3t}$.  \autoref{fig:MC_TVAR_133_cutoff3_shockto3_T200} shows the results of the exercise. The pink continuous line shows the true linear impulse response associated with regime 1. The squared black line captures the true generalized impulse responses. The two lines differ. The dashed line shows the estimated response from the linear-LP. The figure shows that this response lies between the true generalized and the true linear impulse responses. By contrast, the impulse response estimated via BART-LP better estimates the true generalized impulse response. The model correctly captures the evolution of all three variables. It detects that the endogenous evolution across regimes makes the response of the first two variables more pronounced compared to a linear model that remains in regime 1.

\subsection{Sign-dependent Moving Average model}  

The third model we use for simulation is 
\begin{align}
	\boldsymbol{y}_t &= \sum_{l=0}^{20} \boldsymbol{\beta}_{gdp,~l} \cdot \epsilon_{gdp,t-l} ~+~ \sum_{l=0}^{20} \boldsymbol{\beta}_{\pi,l} \cdot  \epsilon_{\pi,t-l} ~+~  \\
	&~~+ \sum_{l=0}^{20} \Big[ \boldsymbol{\beta}_{ff,l}^{+} \cdot \text{I}(\epsilon_{ff,t-l} \ge 0) ~+~ \boldsymbol{\beta}_{ff,l}^{-}  \cdot \text{I}(\epsilon_{ff,t-l} < 0) \Big] \cdot \epsilon_{ff,t-l}, \\
	\boldsymbol{\epsilon}_t &\sim N(\boldsymbol{0}, I),
\end{align} with $\boldsymbol{y}_t = \big(gdp_t, ~\pi_t, ~ff_t\big)'$ a vector containing real GDP, inflation, and the federal funds rate, and $\boldsymbol{\epsilon} = \big(\epsilon_{gdp,t}, ~\epsilon_{\pi,t}, ~\epsilon_{t,ff}\big)'$ a vector of structural shocks. The model is a moving average process of order 20 driven by three structural shocks, a GDP shock, an inflation shock, and a monetary policy shock. The monetary policy shock affects the variables differently at each horizon $t+h$, depending on whether the monetary policy shock at time $t$ is positive or negative. The true impulse responses are captured by $\{\boldsymbol{\beta}_{gdp,~l}\}_{l=0}^{20}$ for the GDP shock, by  $\{\boldsymbol{\beta}_{\pi,~l}\}_{l=0}^{20}$ for the inflation shock, and by $\{\boldsymbol{\beta}_{ff,~l}^+,~\boldsymbol{\beta}_{ff,~l}^-\}_{l=0}^{20}$ for the positive and negative monetary policy shock.

We calibrate the model following an approach similar to \cite{Barnichon2019}. We first estimate a recursive VAR model on US real GDP, inflation and the federal funds rate to estimate linear impulse responses to a GDP shock, an inflation shock and a monetary policy shock. We then set $ \{ \boldsymbol{\beta}_{gdp,l},~ \boldsymbol{\beta}_{\pi,l} , ~\boldsymbol{\beta}_{ff,l}^{-}\}_{l=0}^{20}$  equal to the estimated impulse responses to three shock: the GDP shock, the inflation shock, and the monetary policy shock. Last, we set $\boldsymbol{\beta}_{ff,l}^{+}=\boldsymbol{\beta}_{ff,l}^{-}$ for every $l$, with two exceptions: (1) at horizons $l = 2,3$ the first entry of $\boldsymbol{\beta}_{ff,l}^{+}$ equals 3 times the first entry $\boldsymbol{\beta}_{ff,l}^{-}$, and (2) at horizons $l=7,8,..,20$ the second entry of $\boldsymbol{\beta}_{ff,l}^{+}$ equal 3 times the second entry of $\boldsymbol{\beta}_{ff,l}^{-}$. This implies a data generating process in which positive monetary policy shocks affect output and inflation 3 times more than negative shocks in the short term (for output) and in the medium term (for inflation) horizons.

\begin{figure}[hhh!!!]      
	\centering 
	\caption{Monte Carlo simulation - Sign-dependent Moving Average} 
	\includegraphics[viewport = 110 320 490 525, scale = 0.93]{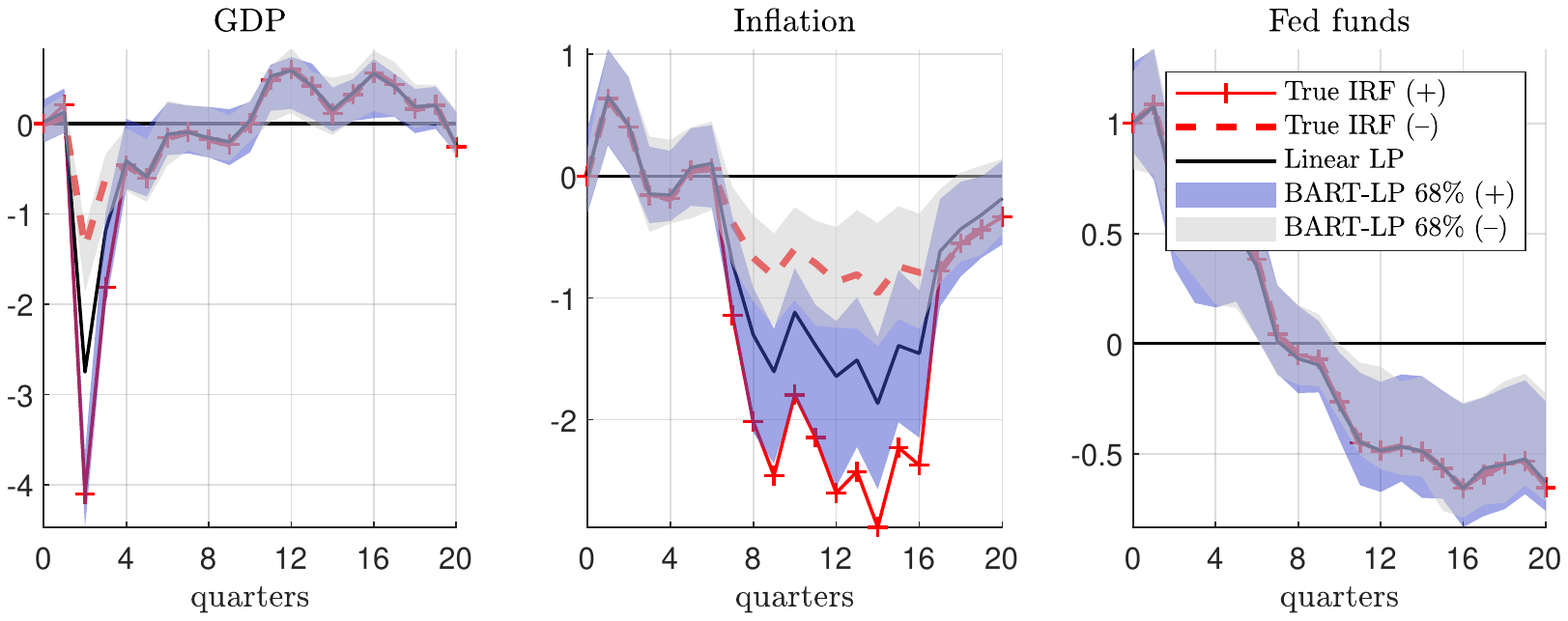} 
	\label{fig:MC_SDMA_45_T200}    
\end{figure}

We design the simulation exercise as follows. We generate an artificial dataset of size $T=400$, discard the first 100 observations and use the remaining 300 for estimation. We estimate the BART-LP model by controlling for the true shocks driving the data, together with up to 2 lags of the shocks and the endogenous variables. Last, we replicate the analysis 100 times and store the pointwise responses from the linear and the BART LPs, computing separately the generalized impulse responses associated with $d\pm 1$.  We set $J=250$.

\autoref{fig:MC_SDMA_45_T200} shows the impulse responses. The $(+)$ and $(-)$ red lines show the true impulse responses associated with a positive and a negative monetary policy shock, respectively. The responses to a negative shock are reported with flipped sign to improve the comparison. By construction, the true response of the federal funds rate does not change in the sign of the shock, while the true response of GDP and inflation is 3 times stronger in response to a positive shock in periods in which the estimated linear responses imply strong responses of GDP and inflation. The black continuous line shows the pointwise median response associated with the linear LP. Note that it sits approximately halfway through the true positive and negative responses. The blue and grey shaded areas report the 68\% pointwise posterior bands associated with the BART-LP model following a positive and a negative shock, respectively. The BART-LP methodology estimates the true response of GDP correctly both qualitatively and quantitatively. It correctly detects that positive shocks have stronger short run effects on GDP than negative shocks, and also estimates the true values of the responses accurately. As for inflation, BART-LP correctly estimates the differential response qualitatively, detecting the stronger effect of positive shocks. While the response to a positive shock is somewhat underestimated, the response to a negative shock sits approximately in the middle of the estimated bands. All in the all, the model correctly detects that positive and negative shocks do not have the same effect.

\begin{table}[hhh!!] 
	\footnotesize
	\centering
	\caption{Monte Carlo Probability that the effect is stronger for a positive shock}
	\begin{tabular}{r|ccc|ccc|ccc}
		\hline
		& \multicolumn{3}{c}{GDP} & \multicolumn{3}{c}{Inflation} & \multicolumn{3}{c}{Fed funds} \\
		$h$ & \scriptsize $T=100$   & \scriptsize $T=200$    & \scriptsize $T=300$    & \scriptsize $T=100$   & \scriptsize $T=200$    & \scriptsize $T=300$    & \scriptsize $T=100$   & \scriptsize $T=200$    & \scriptsize $T=300$  \\
		0     & 49    & 48    & 54    & 54    & 52    & 59    & 58    & 47    & 56 \\
		2     & \textbf{100}   & \textbf{99}    & \textbf{100}   & 59    & 60    & 54    & 56    & 44    & 46 \\
		3     & \textbf{81}    & \textbf{86}    & \textbf{91}    & 50    & 53    & 47    & 54    & 39    & 46 \\
		6     & 54    & 43    & 38    & 55    & 49    & 49    & 47    & 44    & 49 \\
		14    & 49    & 51    & 45    & \textbf{73}    & \textbf{93}    & \textbf{89}    & 43    & 52    & 53 \\
		15    & 50    & 54    & 40    & \textbf{68}    & \textbf{81}    & \textbf{82}    & 49    & 48    & 49 \\
		\hline 
	\end{tabular}%
	\label{tab:SDMAmodel}%
\end{table}%

We further assess the ability of the model to detect differences in positive and negative monetary policy shocks by replicating the analysis over datasets of different lengths. We set the estimation sample period equal to $T=100, 200, 300$ observations, keeping the initial discarded observations to 100. For each sample size, we replicate the analysis 100 times and compute the percentage of Monte Carlo iterations in which the effect is stronger after a positive rather than a negative shock in absolute value. \autoref{tab:SDMAmodel} reports the results for few illustrative horizons, indicating in bold the horizons and variables for which the data generating process features a non-linearity in the response to positive and negative shocks. On impact, the data generating process implies no non-linearity between positive and negative shocks. Indeed, the estimated impulse responses are well distributed across iterations, with around 50\% of the iterations detecting a stronger effect of positive shocks and the remaining 50\% detecting the opposite. At horizons 2-3, close to all iterations detect that the effects of a monetary shock are stronger on GDP, consistent with the data generating process, while correctly not finding evidence of non-linearities in the response of inflation and the policy rate. 6 horizons from the shocks the model correctly detects no non-linearity in any variable. At horizons 14-15 it correctly detects a stronger effect on inflation associated with positive compared to negative shocks, and no non-linearity in the remaining variables.

\section{Empirical analysis} \label{empirics}

In this section, we apply the proposed model to two recent issues that have featured prominently in the empirical literature on non-linear macroeconomic dynamics.

\subsection{Fiscal shocks during recessions and expansions}

In a seminal contribution \cite{auerbach2013fiscal} use a smooth transition VAR model to show that the response of output to government spending shocks is larger during recessions. However, this evidence was disputed by \cite{Ramey2018} who use a 120 year sample of quarterly data to estimate the response to military spending news shocks and show that there is no systematic difference between the spending multiplier in normal periods and those characterised by slack. Both papers use a non-linear LP model, but postulate a functional form for the transition, using either a logistic function or an indicator function.

We build on section VI-B of  \cite{Ramey2018}  and re-visit the analysis of state-dependent fiscal multipliers using our flexible BART-LP model from equation \eqref{FLP}. We set $y_{t+h}$ equal to $h$-period ahead of (1) military spending news $(news_{t})$, (2) real per-capita government spending $(g_{t})$, and (3) real per-capita GDP $(y_{t}) $.\footnote{The data set collated by \cite{Ramey2018} is quarterly and runs from 1889 Q3 to 2015 Q3. The variables used in our analysis are downloaded from the website of the \href{https://www.journals.uchicago.edu/doi/suppl/10.1086/696277}{Journal of Political Economy}. Government spending and GDP are transformed via the procedure described in \cite{Gordon2010TheEnd}. } We first estimate a linear SVAR model in these three variables and estimated the impulse vector $\boldsymbol{d}$ associated with the first shock. We then compute impulse responses as explained in \autoref{sec_irfs}. We include 4 lags of the variables into the model. We set the number of trees to 250 and use 2,000 posterior draws, with a burn-in of 1,000 draws.

We study non-linear effects of government spending shocks as follows. We define periods of recessions as periods where $y_{t}$ is less than its 20th percentile, while expansions denote periods where $y_{t}$ is above the 80th percentile. We then compute generalized impulse responses by generalizing over randomly drawn values of the conditioning values. Compared to \cite{auerbach2013fiscal}  and \cite{Ramey2018}, our framework also allows for the computation of impulse responses that potentially differ for the size of the shock. For this reason, we compute the generalized impulse responses to either a contractionary or an expansionary fiscal spending shock, both studied either in a recession or an expansion.

\begin{figure}[hh!!]
	\centering
	\caption{Response to positive (top panel) and
		negative (bottom panel) military news shocks during recessions and expansions \label{Application_Fiscalpolicy_1}} 
	\includegraphics[viewport = 110 245 490 580, scale = 1]{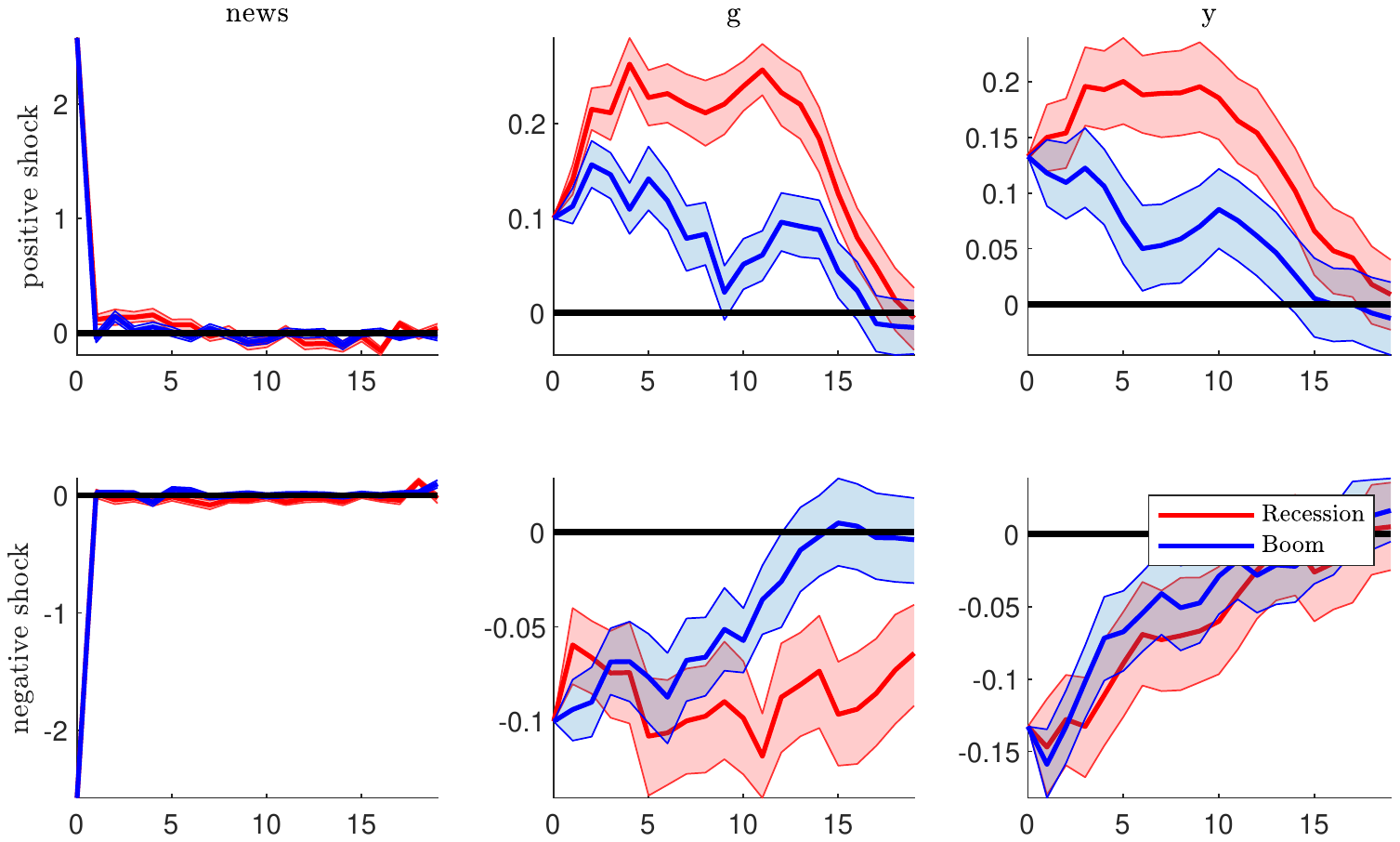} 
\end{figure}

\begin{figure}[hh!!] 
	\centering
	\caption{Cumulated multiplier \label{Application_Fiscalpolicy_2}} 
	\includegraphics[viewport = 110 295 490 530, scale = 1]{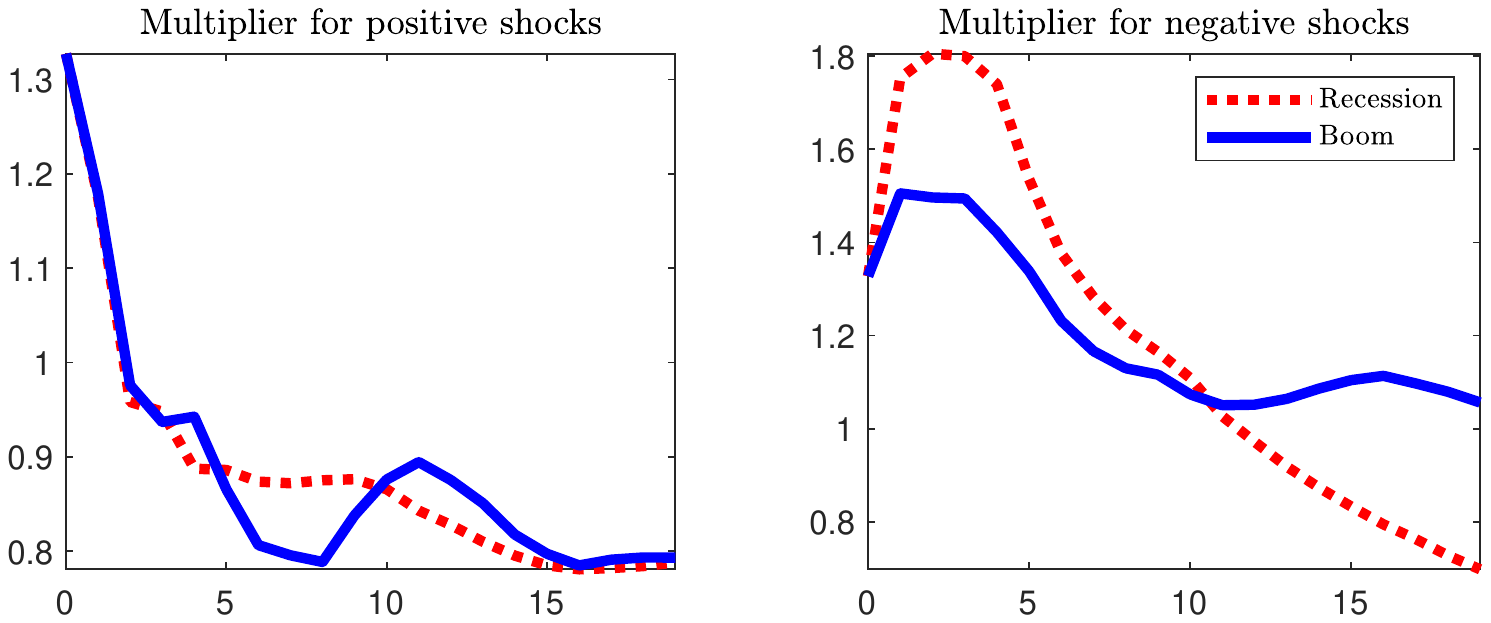} 
\end{figure}

The top panel of \autoref{Application_Fiscalpolicy_1} presents the response to an expansionary spending shock normalised to increase $g$ by 0.1 units on impact. The response of $g$ and $y$ is estimated to be more persistent during recessions. However, the response of  these two variables moves closely together in the two regimes. As a consequence, the spending multiplier (defined as the ratio of the cumulated response of $y$ to the cumulated response of $g$) is similar across regimes. This can be seen from the left panel of \autoref{Application_Fiscalpolicy_2}, which displays  the estimated multipliers and shows that evidence for a systematically  larger multiplier during recessions is weak. However, it is interesting to  note that contractionary spending shocks produce different dynamics. The  bottom panel of \autoref{Application_Fiscalpolicy_1} shows that the response of $y$ to these shocks during expansions is less persistent. As a consequence the multiplier for spending  cuts during these periods is systematically smaller than the estimate during recessions. This distinction helps reconcile the different results found by \cite{auerbach2013fiscal} and \cite{Ramey2018} by documenting that state-specific non-linearities can be different depending on the sign if the shock.

	\subsection{Non-linear impact of Financial shocks}

In a recent contribution \cite{forni2022nonlinear} use a non-linear vector moving average model to show that large negative financial shocks have a proportionally larger impact on the US economy. We revisit this question using the flexible local projection proposed in this paper.  We use the model from equation \eqref{FLP} setting $y_{t+h}$ equal to (1) growth of industrial  production $( IP_{t}) $, (2) CPI inflation $(CPI_{t}) $, (3) unemployment rate $( U_{t}) $, (4) Excess bond premium $( EBP_{t}) $ of \cite{gilchrist2012credit}, (5) stock returns $( STOCK_{t})$ and (6) federal funds rate $(FFR_{t}) $. The data is monthly and the sample runs from 1973M1 to  2022 M2.\footnote{The excess bond premium is downloaded from the federal reserve \href{https://www.federalreserve.gov/econres/notes/feds-notes/updating-the-recession-risk-and-the-excess-bond-premium-20161006.htm}{website}. Stock returns are calculated using the Standards and Poor total return index obtained from Global Financial database. The remaining variables are taken from the FRED database.}

We follow the identification approach by \cite{forni2022nonlinear}  and identify the monetary policy shock as the shock that affects fast moving variables but not slow moving variables. We estimate a preliminary linear VAR model using the above ordering of the variables and estimate the impact vector associated with the entry of the excess bond premium in a recursive identification. We then add 6 lags of all variables. We simulate positive and negative shocks, considering either a shock that moves the excess bond premium by $\pm$50 basis points, or by $\pm$300 basis points. 
	
		\begin{figure}[hh!!]  
		\centering
		\caption{Response to contractionary and
			expansionary financial shocks that change the excess bond premium by 50
			basis points. The response to expansionary shocks has been multiplied by -1
			for the purpose of comparison.  \label{Application_Financial_3}} 
		\includegraphics[viewport = 110 245 490 580, scale = 1]{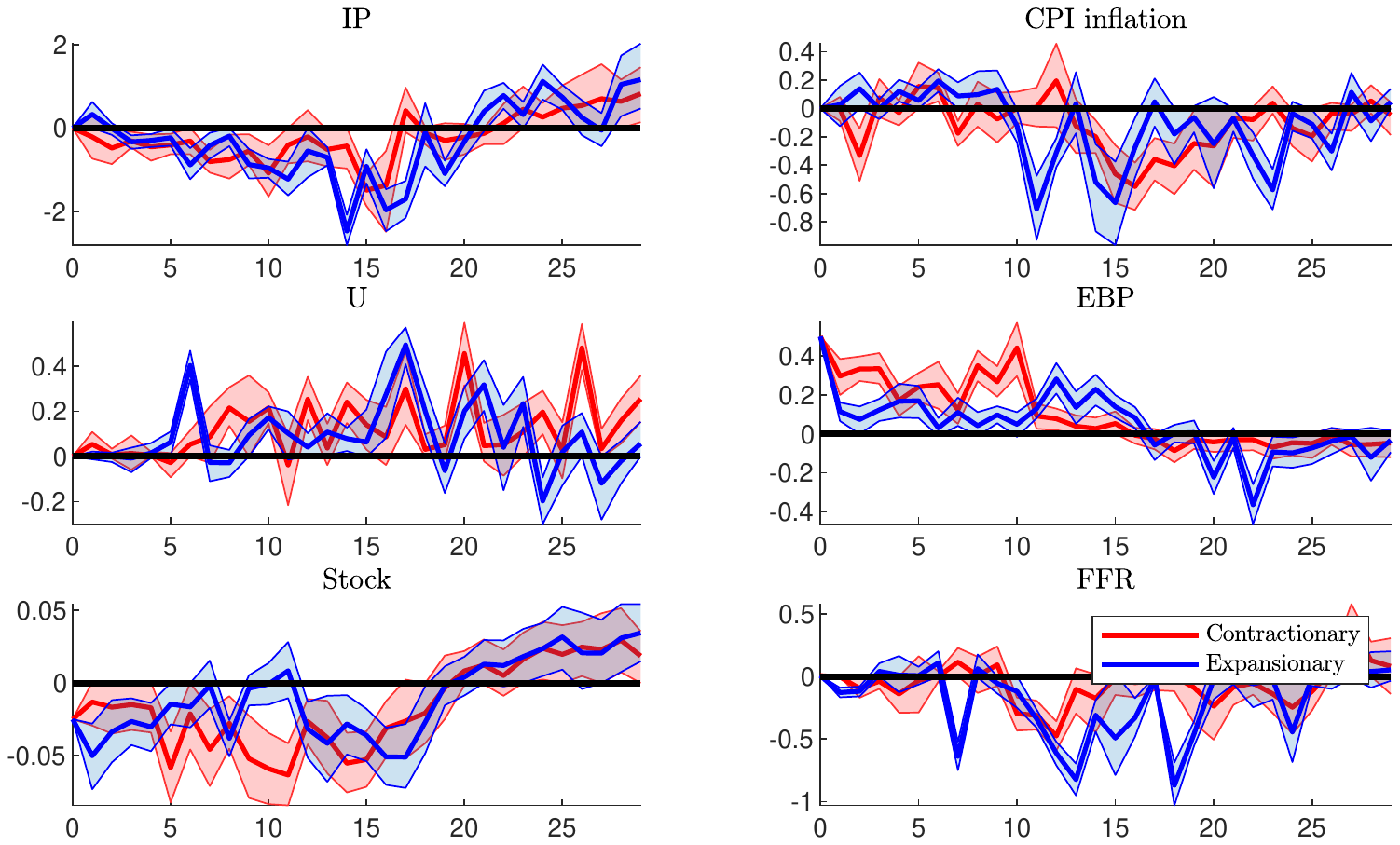} 
	\end{figure} 
	
	\begin{figure}[hh!!] 
		\centering
		\caption{Response to contractionary and
			expansionary financial shocks that change the excess bond premium by 300
			basis points. The response to expansionary shocks has been multiplied by -1
			for the purpose of comparison \label{Application_Financial_4}} 
		\includegraphics[viewport = 110 245 490 580, scale = 1]{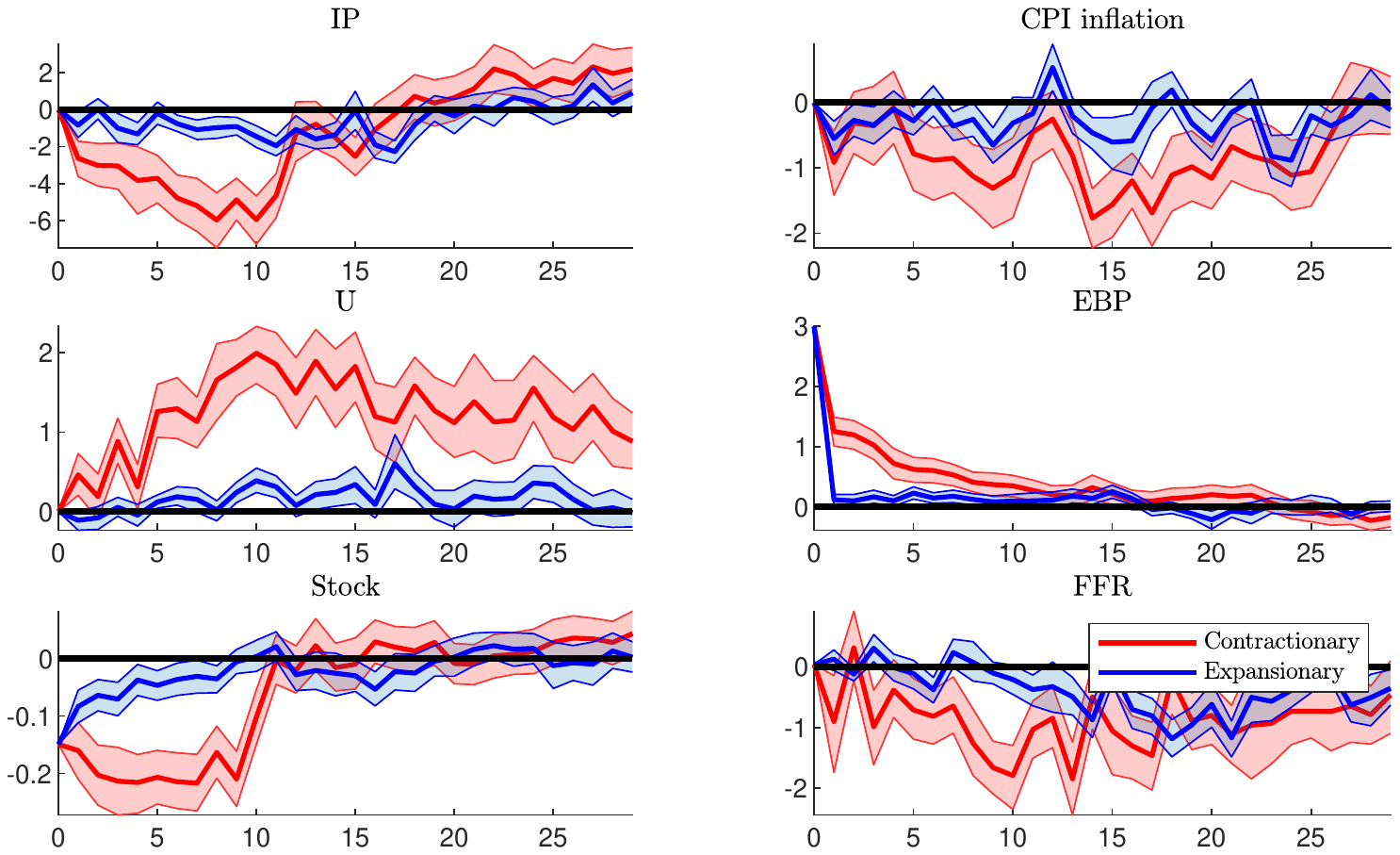} 
	\end{figure}   
	
	The results of the analysis are shown in \autoref{Application_Financial_3} and \autoref{Application_Financial_4}, which report the impulse responses associated with the small or the large exogenous variation of the excess bond premium, respectively. The response to expansionary shocks has been multiplied by -1, to improve the comparison. \autoref{Application_Financial_3} shows that when the magnitude of the financial shock is relatively small, systematic evidence for differences in responses across the sign of the shock is largely absent. However, for large shocks there is clear evidence  of sign non-linearity. \autoref{Application_Financial_4} shows that shocks that increase $EBP_{t}$ by 300 basis points are associated with large declines in output, inflation, stock returns and the interest rate while the unemployment rate rises. The responses to expansionary shocks of the same size are substantially smaller in magnitude. These results broadly confirm the findings reported by \cite{forni2022nonlinear}.

	\section{Conclusions} \label{sec_conclusions}

Local projections are widely used in Macroeconometrics, as they provide a flexible tool to estimate impulse responses to structural shocks of interest. However, the most popular linear specification of local projections introduces the assumption of a linear relationship among variables within each horizon $h$ considered. This paper introduces a flexible local projection that generalises the model of \cite{Jorda2005} to a non-parametric setting by using Bayesian Additive Regression Trees (BART). Using Monte Carlo experiments, we show that the model is able to capture impulse response non-linearities driven by state-dependence, sign-dependence or size-dependence. 

We apply our methodology to US fiscal and financial shocks. We show that while it is true that the fiscal multiplier is stronger in recession, as advocated by  \cite{auerbach2013fiscal}, this holds true only in response to a contractionary fiscal shock. In response to an expansionary shock, we confirm the result by \cite{Ramey2018}, namely that the fiscal multiplier does not change significantly between recession and expansion. We then confirm the results by \cite{forni2022nonlinear} that financial shocks have non-linear effects on the economy. A financial shock that increases the cost of borrowing generates contractionary effects that increase more than proportionately in the size of the shock. This suggests that strong negative financial shocks generate stronger effects that would be otherwise predicted using linear models.

\newpage 
\clearpage

\addcontentsline{toc}{section}{References}
\bibliographystyle{agsm}
\bibliography{references_new} 
\clearpage

\end{document}